# Quantum Locality?

Henry P. Stapp

*Theoretical Physics Group*
*Lawrence Berkeley National Laboratory*
*University of California Berkeley, California 94720*

**Abstract.** Robert Griffiths has recently addressed, within the framework of a 'consistent quantum theory' that he has developed, the issue of whether, as is often claimed, quantum mechanics entails a need for faster-than-light transfers of information over long distances. He argues that the putative proofs of this property that involve hidden variables include in their premises some essentially classical-physics-type assumptions that are fundamentally incompatible with the precepts of quantum physics. One cannot logically prove properties of a system by establishing, instead, properties of a system modified by adding properties alien to the original system. Hence Griffiths' rejection of hidden-variable-based proofs is logically warranted. Griffiths mentions the existence of a certain alternative proof that does not involve hidden variables, and that uses only macroscopically described observable properties. He notes that he had examined in his book proofs of this general kind, and concluded that they provide no evidence for nonlocal influences. But he did not examine the particular proof that he cites. An examination of that particular proof by the method specified by his 'consistent quantum theory' shows that the cited proof is valid within that restrictive version of quantum theory. An added section responds to Griffiths' reply, which cites general possibilities of ambiguities that make what is to be proved ill-defined, and hence render the pertinent 'consistent framework' ill defined. But the vagaries that he cites do not upset the proof in question, which, both by its physical formulation and by explicit identification, specify the framework to be used. Griffiths confirms the validity of the proof insofar as that framework is used. The section also shows, in response to Griffiths' challenge, why a putative proof of locality that he has described is flawed.

# INTRODUCTION

Robert Griffiths begins his recent paper *Quantum Locality* [1] with the observation that "The opinion is widespread that quantum mechanics is nonlocal in the sense that it implies the existence of long range influences which act instantaneously over long distances, in apparent contradiction to special relativity". He says that the purpose of his paper "is to move beyond previous discussions by employing a fully consistent quantum mechanical approach" to "argue that the supposed nonlocal influences do *not* exist" and to "establish on the basis of quantum principles a strong statement of quantum *locality*: the objective properties of an isolated individual (quantum) system do not change when something is done to another non-interacting system."

Griffiths' claims, if valid, would constitute an extremely important achievement: it is difficult to find an issue as central to our understanding of nature as the question of whether or not far-flung parts of the universe are tied together by long-range faster-than-light transfers of information .

Almost all of Griffiths' paper is directed against arguments for nonlocality that are based on the concept of hidden variables: the paper is directed primarily against arguments that have stemmed directly from the works of John Bell pertaining to local deterministic and local stochastic *hidden-variable* theories. However, the local *stochastic* hidden-variable theories have been shown by Stapp [2], and also by Fine[3], to be essentially equivalent to local *deterministic* hidden-variable theories. But these latter theories are theories of an essentially classical-physics type, with statistically distributed unobservable hidden variables. Such theories could include Bohm's pilot-wave model if it were stripped of its nonlocal-interaction feature, which is, however, essential to its structure and its success, particularly in applications to the EPR-type correlation experiments that are the basis of the arguments for nonlocal influences.

In view of this basically *classical* character of the hidden-variable theories, it is obviously going to be extremely difficult to deduce, in any logically sound way, the properties of a quantum-mechanical world from the properties of hidden-variable models: How can one pass, logically, from fact that one needs to add nonlocal influences to any essentially classical model, in order to fit the quantum predictions, to conclusions about the quantum mechanical universe itself? The logical difficulty in deriving such a conclusion is that the hidden-variable premises contain classical reality assumptions that are incompatible with basic quantum concepts. In view of this basic logical problem, it is clear that a search for a strictly rational proof of the existence within the quantum universe of nonlocal influences should focus on arguments that do not use hidden variables; arguments that are not based on the failure of local hidden-variable theories! Griffiths nevertheless confines his attention mainly to arguments for nonlocality based on the failure of local hidden-variable theories.

Commenting upon this severe curtailment of the scope of his arguments Griffiths laments that "In an argument of modest length it is impossible to deal with all the published arguments that quantum theory is beset with nonlocal influences… In particular we do not deal with …Stapp's counterfactual arguments.  …the problems associated with importing counterfactual reasoning into the quantum domain are treated in some detail in Ch. 19  of [4], and the conclusion is the same: there is no evidence for them."

In this paper I shall show that the methods that Griffiths developed lead, rather, to the opposite conclusion. His "fully consistent quantum approach" *validates* the counterfactual argument that he cites, but does not analyze. The validated nonlocal influence required by the assumed validity of certain predictions of quantum theory is fully concordant with the basic principles of relativistic quantum field theory, which ensure that the phenomena covered by the theory can neither reveal a preferred frame associated with these influences, nor allow "signals" (sender-controlled information) to propagate faster than the speed of light.

## COUNTERFACTUALS IN PHYSICS

The word "counterfactual" engenders in the minds of minds of most physicists a feeling of deep suspicion. This wariness is appropriate because counterfactuals, misused, can lead to all sorts of nonsense. On the other hand, all arguments for the need, in a universe in which the predictions of quantum mechanics hold, for some faster-than-light transfer of information requires considering in a single logical analysis the predictions of quantum theory associated with (at least) four *alternative* possible measurements. Probably the only logically sound way to do this, without bringing in hidden-variables, is to use counterfactuals. This can be done in a completely logical and rational way. Indeed, Griffiths takes pains to show how valid counterfactual reasoning is to be pursued and validated within his "consistent quantum theory". His conclusion pertaining to the validation of counterfactual reasoning is the basis of the present work.

Griffiths begins his discussion of counterfactuals [4, p. 262] by noting that "Unfortunately, philosophers and logicians have yet to reach agreement about what constitutes valid counterfactual reasoning in the classical domain." It is certainly true that philosophers fall into disputes when trying to formulate general rules that cover all of the conceivable counterfactual situations that they can imagine, in a classical-physics, and hence deterministic, setting. But such a setting is strictly incompatible with the notion of "free choices" that underlies the idea of alternative possibilities. But what will be examined here is only a very simple special case, one in which the quantum mechanical laws (predictions) *themselves* specify all that we need to know about the outcomes of the contemplated measurements, and in which alternatives arising from alternative possible choices become theoretically possible because of the allowed entry of elements of chance into the dynamics of the choices of which measurements will be performed.

As a brief introduction to the subject of counterfactual statements, consider the following simple classical example: Suppose an electron that is moving in some fixed direction with definite but unknown speed is shot into a region in which there is an electric field E that is known to be uniform at one or the other of two known values, E1 or E2, with E2 twice E1. And suppose two detectors, D1 and D2, are placed so that one can assert, on the basis of the known laws of classical electromagnetism, that "If E is E1 and detector D1 clicks, then if, *instead*, E is E2, the detector D2 would have clicked." Under the appropriate physical conditions this can be a valid theoretical assertion, even though it cannot be empirically verified, since one can not actually perform both of the contemplated alternative possible experiments. But the postulated physical laws allow one to infer from knowledge of what happens in a certain performed experiment what would have happened if, instead, an alternative possible measurement had been performed, all else being the same. The concept "if, *instead*," becomes pertinent in a quantum context in which this choice between E1 and E2 is controlled by whether a certain quantum detection device "clicks" or not. This choice of which measurement is performed is then not determined by the quantum mechanical laws, but enters as a "random" variable.

Consider in this light the following formulation of a putative argument for the need for faster-than-light transmission of information.

Suppose in each of two space-like separated regions, L and R , with L earlier than R (in some frame) there will be performed one or the other of two alternative possible measurements, with each measurement having two alternative possible outcomes. The choices between alternative possible measurements are to be specified in way that can be considered, within the quantum framework, to be "free choices": they are not specified by any known law or rule. The question at issue is whether, under these conditions, it is possible to satisfy the orthodox predictions of quantum mechanics in the four alternative possible measurement situations, without allowing information about the free choice made in either region to be present in the other region.

Notice that the only things that enter the argument are the random choices of which macroscopically described measurement is performed in each region, and the predictions of the theory about which macroscopically described outcomes then appear. No microscopic quantities or properties enter into the argument.

## GRIFFITHS' CONSISTENT QUANTUM THEORY

The proof in question of the need for faster-than-light transfer of information was given in [5], and repeated in the last two pages of [6]. But the purpose of this paper is not to recall old results. It is rather to comment upon Griffiths' "consistent quantum theory" approach, which has attracted interest due to references to it by Murray Gell-Mann and Jim Hartle (who, in contrast to Griffiths, use it in a "Many-Worlds" context), and in particular to show that the counterfactual argument cited but not analyzed by Griffiths is, contrary to Griffiths' implicit claim, *validated* within his "consistent quantum theory" framework, as currently defined. This validation of the need for faster-than-light transmission of information within the "consistent quantum theory" framework constitutes a serious failing of that approach, insofar as it claims to be superior to the von Neumann approach because it does not lead to nonlocal influences.

I begin by describing Griffiths' general theory and its relationship to the orthodox quantum theory of von Neumann, to which it is contrasted.
,
"Measurements" play a very important role in orthodox quantum mechanics. But they are not generated by the quantum evolution in accordance with the Schroedinger equation. The physical act of performing a measurement on a quantum system and getting a positive empirical outcome is represented in the orthodox quantum mathematics by the action a corresponding *projection operator* on the prior quantum state.

Generalizing from the concept of a set alternative possible measurement *outcomes* at one single time one arrives at the concept of a "framework", which involving a sequence times $\{t_0, t_1, t_2, …, t_f\}$, with $t_{i+1} > t_i$ and for each of these times $t_i$ a set of orthogonal projection operators that sum to unity .

A "history" is a time-ordered set of (Heisenberg Picture) projection operators (all operating in the usual Hilbert space of the full physical system) with one projection operator selected from the set at each time $t_i$ . The different alternative possible "histories" labeled by index k (which runs over the set of possible histories) are mapped (by Griffiths' chain operator) into operators represented by the symbols $F_k$. For each $F_k$ the Hermitian conjugate of $F_k$ is represented by $G_k$. Let "rho" represent the initial density matrix. Then the set of histories is called a "consistent" if and only if Trace ($G_g$ rho $F_k$) is zero when g is different from k. This condition is automatically satisfied if, as in the case to be examined here, all of the occurring projection operators, in context, commute. In our case, every nonzero $F_k$ can be identified by a trajectory that moves from left to right on a temporal tree graph that starts from a single line on the far left, and ends at one of sixteen possible lines on the far right, with each non-final segment of the tree graph having a binary branching into two lines at its right-hand endpoint, which occurs at one of the four times $t_i$ at which a choice (of a measurement or an outcome) is made. This leads to sixteen possible lines on the far right of the tree graph. Purely for simplicity, one can take the evolution between measurements to be represented by the unit operator. In order to allow an easy graphical check on Griffiths' rules for validating counterfactual arguments one can, and should, prune away any branches that are required to have zero amplitude for the Hardy initial state.

Griffiths' procedure for checking the validity of counterfactual reasoning is to draw a tree graph that starts at the far left with a single horizontal line that represents the original (in our case, Hardy) state. In our case this line bifurcates at time $t_1$ into an upper branch labeled by ML1, and a lower branch labeled by ML2. These two branches represent the two alternative possible observer-selected settings of the device in the earlier region L. Then at time $t_2$ the line ML1 bifurcates into an upper branch labeled by ML1+, and a lower branch labeled by ML1-, and the branch ML2 bifurcates in similar way into ML2+ and ML2-. These branches represent the two alternative possible states of the outcome indicator (pointer) on device ML set at state of readiness ML1, and, alternatively, on the device ML set at state of readiness ML2. At time $t_3$, each of these four branches bifurcates into an upper branch MR1 and a lower branch MR2, and then at time $t_4$ each of the eight branches bifurcates into a plus and a minus branch, giving one branch for each of the sixteen orthogonal states of the pair of apparatuses together with their respective pointers. This graph represents one single framework, within which the entire argument can be carried out, thereby satisfying Griffiths' crucial "single framework rule". Due to the orthogonality of the states representing the alternative possible device settings and of the alternative possible pointer locations in each region, and the orthogonality of the apparatus-pointer "outcome" states in the two regions L and R, Griffiths' condition of "consistent histories" is satisfied. Thus we can proceed to check Griffiths' condition for valid counterfactual reasoning.

The pertinent counterfactual statement has the form:

SR: "If MR1 is performed and the outcome MR1+ appears, then if, instead of MR1, rather MR2 is performed then the outcome MR2+ must appear."

If the initial state is the Hardy state, then Hardy [7] gives four pertinent predictions of quantum theory:

S1: If ML1 and MR1+, then ML1+. [Hardy's (14.a)]
S2: If ML1+ and MR2, then MR2+ [Hardy's (14.c)]
S3: If ML2+ and MR1, then MR1+. [Hardy's (14.b)]

S4: If ML2+ and MR2, then sometimes MR2-."   [Hardy's (14.d)]

[Connection to Hardy's notation:

| Hardy's | Stapp's |
|---|---|
| $U_1 = 0$ | ML1+ |
| $U_1 = 1$ | ML1- |
| $D_1 = 0$ | ML2- |
| $D_1 = 1$ | ML2+ |
| $U_2 = 0$ | MR1- |
| $U_2 = 1$ | MR1+ |
| $D_2 = 0$ | MR2+ |
| $D_2 = 1$ | MR2- |

Statement S1 follows from Hardy's (14.a), which entails that, in the Hardy state, if ML1 and MR1 are performed and outcome MR1+ ($U_2 = 1$) appears, then outcome ML1+($U_1 = 0$) must appear---since ML1- ($U_1 = 1$) cannot appear. Statement S2 follows from (14.c), [If MR2 and ML1 are performed and MR2 has outcome -, then ML1 must have outcome -: Use the fact that A→B is equivalent to NotB→NotA. Statement S3 is a direct translation of Hardy's (14.b), and S4 follows from Hardy's (14.d), which asserts that the probabability that both ML2+ ($D_1 = 1$) and MR2- ($D_2 = 1$) appear is (with nonzero A and B) nonzero.]

It is a straightforward exercise to show that if the initial state is the Hardy initial state, and if it is assumed that an outcome that occurs and is recorded in the *earlier* region L is left unchanged if instead of MR1 rather MR2 is performed *later* in R, then the statement SR is true if ML1 is performed in L but is false if ML2 is performed in L: The truth of the statement SR about possible happenings in R depends upon which experiment is "freely chosen" in the region L, which is spacelike separated from region R

Griffiths' validation of SR in the ML1 case follows from the fact that if the choice in L is ML1 then starting on branch MR1+, the quantum prediction S1 justifies the move back to the "pivot point" where ML1+ branches into MR1 and MR2. Then S2 justifies the move forward to MR2+.

But if the choice of measurement in L had been ML2 then sometimes the outcome ML2+ appears. But under that condition, if MR1 is chosen on the right, then S3 implies that the outcome on the right must be MR1+. But in this case where MR1+ must appear, if, instead, MR2 is chosen in R then, virtue of S4, MR2+ sometimes does not appear, and we have a counter example to what was proved true in the case that ML1 was chosen in L. All parts of the argument are represented in the tree graph that corresponds to a "single framework", in accordance with Griffiths very restrictive "single framework rule".

## RESPONSE TO GRIFFITHS REPLY

Griffiths' CQT is based on his concept of "consistent families of histories", called "frameworks". Within any single framework the usual classical laws of logic hold, but different frameworks can lead to different conclusions. It is therefore absolutely essential to any proposed application of the CQT rules pertaining to counterfactuals that the statement of the theorem: (1), identify a suitable framework that allows the theorem to be proved true: and (2) rule out any framework that allows that theorem to be proved false. Frameworks that do not conform to the conditions of the theorem, and hence do not allow the theorem to be either validated or invalidated are not relevant to the issue of the validity of the theorem.

I shall explain, in this rejoinder, why, in the above proof of nonlocality, I use a proper framework; and why the resulting proof settles the locality issue. The various evasions and ambiguities cited by Griffiths[8], which appear to eviscerate almost completely Griffiths' method of validating counterfactual arguments, do not upset my proof: all pertinent ambiguities are removed by the explicitly stated conditions imposed by the statement of the theorem. The fact that, within a consistent formulation of plane geometry, Pythagoras' Theorem might *not* be provable by spurious methods fraught with ambiguities, does not render that theorem false. Within any consistent formalism one correct proof of an unambiguously formulated general property suffices! No conflicting proof allowed by a consistent formalism can disprove a conclusion established by a valid proof.

One nice thing about Griffiths' reply to my paper is that he not only confirms that my proof is valid in the framework that I use, but he also actually shows the diagrams that immediately confirm almost all of what needs to be proved. In particular, one can see immediately from his Fig.1a, which pertains to the case in which the alternative possibilities in L are $[0]_a$(ML1+) and $[1]_a$(ML1-), that if one performs $Z_b$ (MR1) and gets $Z^-_b$ (MR1+) then if instead of $Z_b$ (MR1) one performs $X_b$ (MR2) then, by tracing back from the *unique* $Z^-_b$ (MR1+), to the pivot point where the choice between $Z_b$ (MR1) and $X_b$ (MR2) is made, and them forward along the MR2 branch, one arrives unambiguously at $X^+_b$ (MR2+). This validates, graphically, the first half of what needs to be proved. It confirms, within CQT, the same conclusion that, by using the normal logical meanings of the occurring words, follows unambiguously from the orthodox predictions of quantum theory, in a Hardy-type experiment.

Most of other half is then proved by looking at Figure 1b, which pertains to the case in which the two alternative possible macroscopic outcomes in L are the states that in my proof I call ML2+ and ML2-. Figure 1b shows that in this case the CQM rules for validating the counter-factual statement SR fail: the final outcome in this second case need not be $X^+_b$ (MR2+).

In fact, as Griffiths notes, more can be proved: If MR1 is performed and MR1+ appears, then if, instead of MR1, rather MR2 is performed, then the probability that $X^+_b$ (MR2+) appears can, by adjusting the parameters of the Hardy state, be made arbitrarily close to zero. Then the outcome will almost never be the required MP2+. Thus what was previously proved to be true in region R, in the case in which the apparatus randomly chosen in L is the one specified by ML1, would be false if the apparatus chosen in L were the one specified by ML2. Thus what is true in R depends on which experiment is (randomly) performed in L. The locality property is thereby proved to be false.

The fact that other frameworks exist that do not conform to the conditions of the theorem, and hence do not allow the theorem to be proved either true or false, does not invalidate what was previously proved. On the other hand, if some CQT framework were to exist that conforms to all the conditions of the theorem, and allows locality to be proved true, rather than false, then the CQT theory would not be a consistent theory.

The proof thus rests upon the fact that there is *some* framework in which SR is true with probability one or false with nonzero probability according to which branch is chosen at time $t_1$, where this choice at $t_1$ corresponds to a random physical choice between two setting of the macroscopic apparatus in

L. This result requires the information about the choice made in L to be present in R, in this particular situation. This consequence of the predictions of quantum theory is sufficient to prove nonlocality. The fact that there are other frameworks that do not yield conclusions about the truth or falsity of the theorem in question is irrelevant to the question of the truth of that theorem.

Griffiths notes that "As the projectors associated with $S_{ax}$ do not commute with those associated with $S_{az}$ the two families in (6) and (7) are incompatible: the results of reasoning using one (family) cannot be combined with the results using the other (family): this is the single framework rule discussed at length in CQT."

This circumstance blocks using in one argument the results from the two Figures 1a and 1b, insofar as we do not included in the analysis the two alternative possible apparati that translate microstates before the measurements into associated non-overlapping macroscopic outcomes.

But the statement of the theorem demands the presence of such apparati: it demands the presence of measuring systems that will produce on the macroscopic scale outcomes that are unambiguously correlated to the associated microstates of the particles entering the apparati. For the two apparati, one can consider a Stern-Gerlach device that can be *rotated* so as "measure" either $S_{ax}$ or $S_{az}$. In the first setting one has, at the macro level, state $Z_a$ which goes to $Z^+_a$ or to $Z^{--}_a$ according to whether the incoming microstate is $S^+_{az}$ or $S^-_{az}$, and similarly for the second setting/apparatus, with Z replaced by X. The four *macro*states $\{Z^+_a, Z^{--}_a, X^+_a, X^{--}_a\}$ are macroscopically well separated. This is my set {ML1+,ML1-,ML2+,ML2-}, which I place at time $t_2$. I put at time $t_1$ the pair {ML1, ML2} where ML1 specifies the macroscopic apparatus that yields outcome ML1+ if the particle entering into has microstate $S^+_{az}$ and yields ML1- if the particle entering into it has microstate $S^-_{az}$, and ML2 specifies the macroscopic apparatus that yields outcome ML2+ if the particle entering into has microstate $S^+_{ax}$ and yields ML1- if the particle entering into it has microstate $S^-_{ax}$. Because of the dynamical independence of these two systems, this arrangement in L gives a consistent family of histories: a single framework within which the entire analysis can be carried out.

As Griffiths says in CQT (p. 231) "A quantum mechanical description of a measurement with particular outcomes must obviously employ a framework in which these outcome are represented by appropriate projectors." Because the statement of my nonlocality theorem explicitly demands the presence of precisely these specific apparati, I am obliged to use these apparati in the analysis and "must obviously employ a framework in which these outcome are represented by appropriate projectors. "

The suggestion made by Griffiths that the counterfactual considerations of Sec. 19.4 of CQT be applied in L is off-target, because the logical structure of the argument is such that in the earlier region L the options are not counterfactual, but are just simple disjunctive alternatives, and the required disjunctive character is justified by the satisfaction of the "consistent histories" conditions. The counterfactual aspect of my proof is restricted to region R.

The explicit formulation my theorem thus entitles me, and requires me, to use the framework that represents the particular apparati that are demanded by the statement of the theorem. Then the proof goes through: the truth of SR depends on which apparatus is (randomly) chosen in L. Hence locality is violated! The fact that there are frameworks that have no impact on the truth or falsity of the theorem is logically irrelevant.

Griffiths claims that "Neither Stapp nor anyone else has yet found a defect in the relatively straightforward (no counterfactuals) demonstration of the principle of Einstein locality given in [2], a principle that directly contradicts Stapp's claim of nonlocality."

Suppose, as Griffiths postulates, that the full systems in each of the two experimental space-time regions, A and B -- with B including the system C that consists of two apparati, a random number generator, and a servo-mechanism that randomly puts in place one apparatus or the other -- do indeed evolve according to locally existing conditions, during the common measurement interval from $t_0$ to $t_n$ , as Griffiths specifies. And suppose there is in each region during that interval, is a single particle from a pair of correlated particles, with the apparatus in each region interacting with the particle located in that region. Thus the time $t_0$ must be a time such that the two measured particles are safely within their respective regions. Otherwise the dynamics will not be separable, as demanded.

Overriding all else are the predictions of quantum theory pertaining to outcomes of the various alternative possible combinations of measurements performed on the correlated particles located in two different space-time experimental regions, A and B.

The conditions imposed in Griffiths' putative proof in no way prevents the situation in region A from depending strongly upon which measurement was performed in region B. Griffiths' Figures 1a and 1b, with B the region L, and A the region R, immediately show that the predictions of quantum mechanics entail that the state of affairs in A depends upon which choice of apparatus is made in region B.

If it be asked how, in strict compliance with Griffiths' stringent locality conditions, the situation in A can depend upon the actions performed in B by C, the simple reply is that the action performed by C on the particle in B influences the detection event in B, which is associated with a changes the state of its partner headed for, but not yet arrived at, the experimental space-time region A.

Griffiths may recoil at such a solution. But, regardless of how nature actually works – a matter upon which opinions differ -- this action-at-a-distance description is compatible with how the *predictions of quantum mechanics are actually constructed by the quantum formalism*: this action is often construed as a backward-in-time action. Hence Griffiths' proof does not work: the conditions do not strictly exclude from region A information about what C does in region B, in a quantum context that, evading ontological prejudices about how nature works, accepts the *predictions* of quantum theory as the solidly valid features of quantum mechanics upon which the proofs of nonlocality build.

Thus Griffiths' proposed proof of locality ignores the essential problem, which is that *the procedure by means of which the predictions of quantum theory are actually calculated involves the spooky action at a distance that Einstein abhorred. That aspect might be completely unreal at some basic level, but it is an integral part of the machinery that produces the predictions upon which all arguments for nonlocality build.* Griffiths' Figures 1a and 1b are perpetual pictorial reminders of fact that the correlations predicted by quantum mechanics entail, under appropriate conditions, that what is measured in one region affects properties in the other. It is of course true that the orthodox quantum rules are such that no "signal" can be transmitted faster than light, and this fact can buttress the opinion that the seeming intrusion of nonlocal effects into dynamics is spurious. But it begs the question to insert that opinion into a claimed proof of locality.

## ACKNOWLEDGEMENT

This work was supported by the Director, Office of Science, Office of High Energy and Nuclear Physics, of the U.S. Department of Energy under contract DE-AC02-05CH11231

## REFERENCES

1. Robert Griffiths, Quantum Locality, *Found. Phys*. 41, 705-733 (2011)
2. Henry P. Stapp, *Epistemological Letters*, June 1979, Colloquium on Bell's Theorem, Assoc. F. Gonseth, Case Postale 1081, Bienne, Switzerland. (1979) Reprinted in http://www-physics.lbl.gov/~stapp/AnEquivalenceTheorem.pdf
3. Arthur Fine, Hidden variables, Joint Probabilities, and Bell Inequalities, *Physical Review Letters*, 48, 291-295, (1982)
4. Robert Griffiths, *Consistent Quantum Theory*, Cambridge Univ. Press, Cambridge UK (2002)
5. Henry P. Stapp, A Bell-type theorem without hidden variables. *American Journal of Physics*, 72, 30-33, (2004)
6. Henry P. Stapp, *Mindful Universe: Quantum Mechanics and the Participating Observer*, Springer, Berlin, Heidelberg, (2007).
7. Lucian Hardy, Nonlocality for Two Particles Without Inequalities for Almost All Entangled States, *Physical Review Letters*, 71, 1665-1668 (1993)
8. Robert B. Griffiths, *Quantum Counterfactuals and Locality.* Following paper.